\documentclass[preprintnumbers,article,amsmath,amssymb,floatfix,10pt,prd,onecolumn,
superscriptaddress,nofootinbib]{revtex4}
\usepackage[colorlinks=true, pdfstartview=FitV, linkcolor=blue, citecolor=red, urlcolor=magenta]{hyperref}
\usepackage{bbm}
\usepackage{amsfonts}
\usepackage{mathrsfs}
\usepackage{latexsym}
\usepackage{epsfig}
\usepackage{epstopdf}
\usepackage{epstopdf}
\usepackage{graphicx}
\usepackage{amssymb}
\usepackage{amsmath}
\usepackage{dcolumn}
\usepackage{bm}
\usepackage{float}
\usepackage{color}
\usepackage{comment}
\usepackage{xcolor}
\begin{document}
\title{Thermal fluctuation, deflection angle and greybody factor of a high-dimensional Schwarzschild black hole in STVG}
\author{Qian Li}
\affiliation{Faculty of Science, Kunming University of Science and Technology, Kunming, Yunnan 650500, China.}
\author{Yu Zhang}
\email{zhangyu\_128@126.com (Corresponding author)} \affiliation{Faculty of Science, Kunming University of Science and Technology, Kunming, Yunnan 650500, China.}
\author{Qi-Quan Li}
\affiliation{Faculty of Science, Kunming University of Science and Technology, Kunming, Yunnan 650500, China.}
\author{Qi Sun}
\affiliation{Faculty of Science, Kunming University of Science and Technology, Kunming, Yunnan 650500, China.}
\begin{abstract}
In this work, we study the thermal fluctuation, deflection angle and greybody factor of the high-dimensional Schwarzschild black hole in  scalar-tensor-vector gravity (STVG). Based on the correction of black hole entropy due to thermal fluctuation, we calculate some  thermodynamic  quantities associated with  the correction of black hole entropy. The influence of the first-order and second-order corrections, spacetime dimensionality and STVG parameters on these  thermodynamics quantities are discussed in detail.  Additionally, by utilizing the Gauss-Bonnet theorem, the deflection angle is obtained in the weak field limit and the effect of two parameters on the results is visualized. Finally, we calculate the bounds on greybody factors of a massless scalar field.
\end{abstract}
\date{\today}
\maketitle
\section{Introduction}

Although Einstein's general relativity is one of the successful and well-established gravitational theories in modern physics, general relativity fails to explain many observational results, such as the present stage of cosmic acceleration \cite{Astier:2012ba}, rotation curves of galaxies \cite{Moffat:2013sja} and some cosmological data \cite{Planck:2015fie}. Moreover, general relativity has inherent deficiencies  in the theory, such as the presence of spacetime singularities. Therefore, the problems of general relativity motivate us to research the alternative gravity theories.  One of the modified gravity theories is the scalar-tensor-vector gravity (STVG) proposed by Moffat \cite{Moffat:2005si}, which is based on the action principle and is  presented by the  metric tensor, three scalar fields  and a massive vector field.  Moffat gave the black hole solution in STVG  in another paper \cite{Moffat:2014aja}. What's more, this modified gravity (MOG), i.e., STVG may be considered an alternative to the dark matter problem, which can be solved by changes in the gravity sector. STVG was able to fit the rotation curves of galaxies \cite{Brownstein:2005zz} without considering dark matter and was showing no difference with solar system observational tests. However, Jamali and his colleagues \cite{Jamali:2017zrh} found that a modified version of the STVG, known as mMOG, cannot be deemed as an alternative to the dark matter problem when new constants are introduced in the kinetic term of the scalar field as its coefficients.
	
	The interest in the physical properties of high-dimensional black holes significantly increases, even though high-dimensional black holes have not been directly observed or experimentally supported in comparison with the four-dimension black hole. This has a lot to do with the development of string theory.  In addition, the theoretical importance of higher dimensional black hole solutions was  introduced by Emparan and Reall \cite{Emparan:2008eg}. Tangherlini \cite{Tangherlini:1963bw}  proposed firstly  the solutions of the Schwarzschild and Reissner-Nordstr$\ddot{\text{o}}$m black holes  in $D$ dimensional spacetime. Later,  Myers et al. obtained the Kerr black hole solution in high dimensional spacetime in Ref. \cite{Myers:1986un}.  Recently, Cai et al. \cite{Cai:2020igv}  derived a  high-dimensional static spherically symmetric Schwarzschild  black hole  in STVG, which is a high dimensional extension of STVG theory, and studied its quasinormal modes of a massless scalar field and black hole shadow. This black hole solution is a  link between Einstein's theory and  STVG theory. Specifically, this black hole degenerates  to Schwarzschild-Tangherlini black hole in Einstein's theory with the coupling constant $\alpha$ being zero.
	
The black hole entropy is proportional to the area of the event horizon of the black hole, known as the Bekenstein-Hawking formula \cite{Bekenstein:1973ur}. The black hole entropy is maximum compared with the objects of the same volume in order to avoid the violation of the second law of black hole thermodynamics.  However, due to thermal fluctuation which leads to the concept of the holographic principle  \cite{Easther:1999gk}, the maximum entropy of black holes may be corrected. The corrected term for maximum entropy is generated by the quantum fluctuations in the spacetime geometry rather than the matter field in the spacetime. For large black holes, quantum fluctuations are negligible. When the size of black hole reduces due to Hawking radiation, however, the quantum fluctuations in the spacetime geometry will increase. Thus, there is a logarithmic correction at leading order in black hole entropy  \cite{Das:2001ic}. Upadhyay investigated the effect of thermal fluctuations on a quasitopological black hole and found the negative correction term result leads to a local instability of black holes \cite{Upadhyay:2017qmv}. The influence of logarithmic corrections on the thermodynamics due to thermal fluctuations for a dilaton black holes in gravity's rainbow has been studied in Ref. \cite{Dehghani:2018qvn}.  There are several works are devoted to studying the thermal fluctuation effects on black hole thermodynamics  \cite{Jawad:2017mwt,Shahzad:2018znu,Sharif:2021vex,Khan:2022zcf,Ama-Tul-Mughani:2022wtg,Chen:2021czh,Upadhyay:2019hyw,Khan:2021tzv}.
	
Hawking believed that black holes are  not completely black objects and can emit radiation, known as Hawking radiation \cite{Hawking:1974rv,Hawking:1975vcx}. This lays an important foundation for understanding the thermodynamics of black holes. The Hawking radiation detected at infinity of the black hole differs by a redshift factor, called as greybody factor, from the authentic radiation detected at the black hole horizon. The greybody factor that derives from the transmission amplitude can provide information related to the quantum nature of the black hole \cite{Barman:2019vst}. There are several methods to calculate the greybody factor such as  the bounds on greybody factors \cite{Boonserm:2008zg,Boonserm:2014fja,Boonserm:2017qcq,Okyay:2021nnh}, the WKB method \cite{Kokkotas:2010zd,Konoplya:2020jgt,Li:2022jda} and the  exact numerical approach \cite{Harris:2003eg,Catalan:2014ama,Abedi:2013xua}. In this paper, we choose the bounds on greybody factor due to the fact that it can provide analytical results for the intermediate frequencies and all angular momentum.
	
When light ray encounters a  dense compact object in its trajectory toward a distant observer, the observer will find the light ray has a deflection angle. That is to say, the compact object bends the light ray, which forms gravitational lensing. So gravitational lensing which can be classified as strong gravitational lensing, weak gravitational lensing and micro gravitational lensing is used as a special astronomical tool to check whether general relativity theory is correct.  Concretely, the strong gravitational lensing is used to calculate the magnification and position of the black hole. The weak gravitational lensing can help us to measure different objects' masses or restrict of the cosmological parameter. In addition,  on the cosmic microwave background aspects, the weak gravitational lensing also has an important effect \cite{Lewis:2006fu,Peloton:2016kbw,Pratten:2016dsm}. At present, strong or weak gravitational lensing of compact objects, such as wormholes, black holes and cosmic strings has been widely considered \cite{Chen:2015cpa,Chen:2016hil,Wang:2016paq,Lu:2016gsf,Zhao:2016kft,Zhao:2017cwk,Zhang:2017vap,Abbas:2019olp,Bergliaffa:2020ivp,Wang:2019cuf,Kumaran:2019qqp,Javed:2020frq,Kumar:2020sag,ElMoumni:2020wrf,Javed:2020pyz,Xu:2021rld,Javed:2021arr,Gao:2021luq,Javed:2020lsg}. Part of the work in the above literature is based on Gauss-Bonnet theorem to calculate the deflection angle for the weak gravitational lensing. The Gauss-Bonnet theorem proposed by Gibbon and Werner \cite{Gibbons:2008rj} in 2008, is used to derive the deflection angle for the first time in the context of optical geometry. Since then,  this method has been applied to the weak deflection angles of different black holes \cite{Ishihara:2016vdc,Islam:2020xmy,Zhu:2019ura,Sakalli:2017ewb,Jusufi:2018jof,Ovgun:2018fte,Li:2020wvn,Javed:2020fli,Belhaj:2020rdb}. We will also research the weak  gravitational lensing of a high-dimensional Schwarzschild spacetime in STVG by using Gauss-Bonnet theorem.

Motivated by the above, the purpose of the paper is to study the thermal fluctuation, weak deflection and grey-body factor of the high-dimensional Schwarzschild black hole in STVG. The present paper is structured as follows. In section \ref{sec2}, we briefly introduce a high-dimensional Schwarzschild black hole solution in STVG. Then, we review the physical features of this black hole. In section \ref{sec3}, we study the corrected thermodynamic quantities due to thermal fluctuation. Section \ref{sec4} is devoted to calculating the weak deflection angle using Gauss-Bonnet theorem.  We discuss the bounds on greybody factors in section \ref{sec5}. In the last section, our conclusions are summarized.

Throughout this paper,  the natural system of units ($G_{N}=\hbar=c=1$) is adopted.

	\section{Fundamental spacetime} \label{sec2}
	In the section, we introduce the high-dimensional Schwarzschild spacetime in
	scalar-tensor-vector gravity (STVG) and simply review some thermodynamical properties.  The general action of the STVG theory in D-dimensional spacetime  takes the form \cite{Moffat:2014aja}
	\begin{eqnarray}\label{Q1}
	S_{L}=S_{GR}+S_{\phi}+S_{S}+S_{M},
	\end{eqnarray}
	where
	\begin{equation}
	\label{Q2}
	S_{\rm GR}=\frac{1}{16\pi }\int d^{D}x\sqrt{-g}\frac{1}{G}R, \hfill
	\end{equation}
	\begin{equation}
	\label{Q3}
	S_{\phi}=-\frac{1}{4\pi }\int d^{D}x\sqrt{-g}\left(K-\frac{1}{2}\tilde{\mu}^{2}\phi ^{\mu }\phi _{\mu}\right), \hfill
	\end{equation}
\begin{equation}
\label{Q4}
\begin{aligned}
S_{\mathrm{S}} & =\int d^{D} x \sqrt{-g} \left[\frac{1}{G^{3}}\left(\frac{1}{2} g^{\mu \nu} \nabla_{\mu} G \nabla_{\nu} G-V_{G}(G)\right)  +\frac{1}{\tilde{\mu}^{2} G}\left(\frac{1}{2} g^{\mu \nu} \nabla_{\mu} \tilde{\mu} \nabla_{\nu}  \tilde{\mu}-V_{\tilde{\mu}}(\tilde{\mu})\right)\right],
\end{aligned}
\end{equation}
here  $S_{GR}$ is the Einstein-Hilbert action, $S_{\phi}$ stands for the action of a massive vector field $\phi^{\mu}$, $S_{S}$ denotes the action of the scalar field and $S_{M}$ represents the matter action. The black hole metric in the D-dimensional spacetime has the following form
	\begin{eqnarray} \label{Q5}
	ds^{2}=-f(r)dt^{2}+\frac{dr^{2}}{f(r)}+r^{2}d\Omega ^{2}_{D-2},
	\end{eqnarray}
	with  the  line element $f(r)$ being \cite{Cai:2020igv}
	\begin{equation}\label{Q6}
	f(r)=1-\frac{m}{r^{D-3}}+\frac{Gq^{2}}{r^{2(D-3)}},
	\end{equation}
	where $G$ is the Newton's gravitational constant, $G=G_{N}(1+a)$. And $m$ and $q$ are defined by
	\begin{equation}\label{Q7}
	m\equiv \frac{16\pi GM}{(D-2)\Omega _{D-2}},\qquad q\equiv \frac{8\pi \sqrt{a G_{N}}M}{\sqrt{2(D-2)(D-3)}\Omega _{D-2}},
	\end{equation}
	where the dimensionless parameter $a$ in the form is regarded as a deviation of the STVG  theory from  standard general relativity theory and  $M$ is the black hole mass.  Moreover, $\Omega_{D-2}$ denoting the volume of unit $(D-2)$-dimensional sphere has the form
	\begin{equation}\label{Q8}
	\Omega_{D-2}=\frac{2\pi^{\frac{D-1}{2}}}{\Gamma (\frac{D-1}{2})}.
	\end{equation}
	
	When the dimensionless parameter $a$, we can get a  Schwarzschild-Tangherlini black hole in Einstein's gravity.  Moffat gave a Schwarzschild black hole in STVG  for the case  $D=4$ \cite{Moffat:2014aja}. Moreover,  one can find that there is a similarity between a  high-dimensional Schwarzschild black hole in STVG and a high-dimensional Reissner-Nordstr\"{o}m black hole in Einstein gravity from the metric \cite{Pourhassan:2017kmm}.  The high-dimensional Schwarzschild STVG black hole  possesses up to two horizons
	\begin{equation}
	\label{Q09}
	r_{\pm}=\bigg(\frac{m}{2}\pm\frac{\sqrt{m^{2}-4Gq^{2}}}{2} \bigg)^{2},
	\end{equation}
	where $r_{-}$ and $r_{+}$ represent the Cauchy horizon and  the event horizon, respectively. But Mureika  et al. \cite{Mureika:2015sda} pointed out that the  Schwarzschild black hole in STVG, i.e., MOG black hole, relies only on the mass $M$ and  dimensionless parameter $a$. So $q$ is called  the gravitational charge rather than charge.
	
	The black hole mass in terms of $r_{+}$ has the form
	\begin{equation}
	\label{Q9}
	M=\frac{r_{+}^{D-3} \left(A-\sqrt{A^2-4 G B^2}\right)}{2  G B^2},
	\end{equation}
	where the coefficients A and B are expressed as
	\begin{equation}
	\label{Q10}
	A\equiv \frac{16 \pi G}{(D-2) \Omega_{D-2}},  \qquad    B\equiv \frac{8 \pi  \sqrt{a G_{\rm N}}}{\sqrt{2(D-2) (D-3)} \Omega_{D-2}}.
	\end{equation}
	
	The Hawking temperature is given by
	\begin{equation}
	\label{Q11}
	T_{\rm H}=\frac{1}{4\pi}\frac{df(r)}{dr}|_{r=r_{+}} =\frac{(D-3)\left(A \sqrt{A^2-4 G B^2 }-A^2+4 G B^2 \right)}{8 \pi  G   B^2r_{ +}}.
	\end{equation}
	
	Also, the  Bekenstein-Hawking entropy of this high-dimensional black hole, $S_{0}$,  is  given by
	\begin{equation}
	\label{Q12}
	S_{0}= \frac{\Omega_{D-2} r_{+}^{D-2}}{4}.
	\end{equation}

	\section{Thermal fluctuations}\label{sec3}
	In the section, we investigate the influence of thermal fluctuations on thermodynamic potential of a  high-dimensional  Schwarzschild  black hole  in STVG. First of all, we simply introduce the thermal fluctuation and then  calculate some important modified thermodynamics quantities.
	
	We can not neglect the influence of the thermal  fluctuation on the black hole  thermodynamics when the radius of the black hole decreases and the temperature of the black hole is large. The thermal  fluctuation  will be  regarded as a  perturbation around the state of  equilibrium  if it is small enough.  Using the partition function approach, a general expression for the corrected  entropy area relation is written  as \cite{Pourhassan:2016zzc,Pourhassan:2017rie,Pourhassan:2018wjg,Bubuianu:2018qsq,Sharif:2022ccc,Sharif:2020hid}
	\begin{equation}
	\label{Q14}
	S=S_{0} -\alpha \text{ln}(S_{0}T^{2})+ \frac{\lambda}{S_{0}},
	\end{equation}
	where $\alpha$ is the leading order correction parameter and $\lambda$ is the  second order correction parameter.  The leading order correction is a logarithmic term caused by the thermal fluctuations, and the second order correction  proportional to the inverse to uncorrected entropy is produced by extending the entropy function around the equilibrium.
	
	Using Eqs. (\ref{Q11}) and  (\ref{Q12}), the  corrected entropy of this high-dimension black hole is given as
	\begin{equation}
	\label{Q15}
\begin{aligned}
	S&=\frac{1}{4}r_{+}^{D-2}{\Omega_{D-2}} + \frac{4r_{+}^{D-2}\lambda}{\Omega_{D-2}}-\alpha \text{ln}[\frac{(D-3)^{2}(A^{2}-4GB^{2})(A-\sqrt{A^{2}-4GB^{2}})^{2} r_{+}^{D-4} \Omega_{D-2}}{256G^{2}B^{4} \pi^{2}}].
\end{aligned}
	\end{equation}
	\begin{figure}[t]
		\centering
		\begin{tabular}{cc}
			\includegraphics[width=0.5 \textwidth]{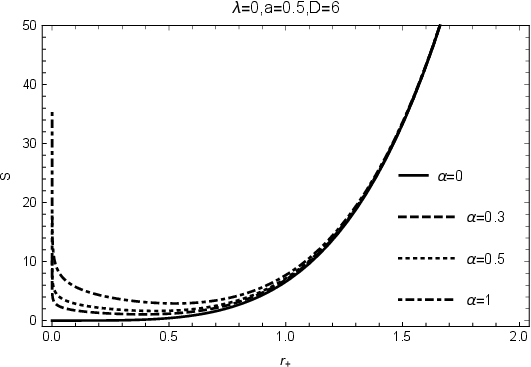}
			\includegraphics[width=0.5 \textwidth]{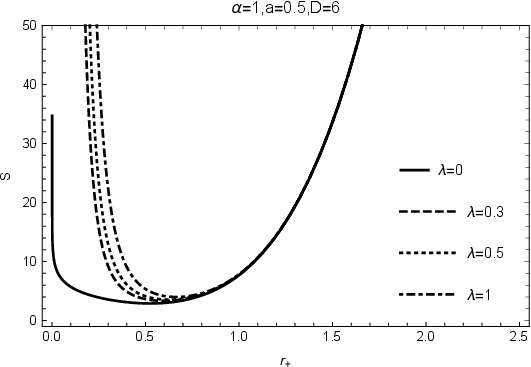}
		\end{tabular}
		\caption{The  entropy $S$ in terms of event horizon $r_{+}$ for different values of $\alpha$ and $\lambda$.}	\label{FIG1}
	\end{figure}
	
	\begin{figure}[t]
		\centering
		\begin{tabular}{cc}
			\includegraphics[width=0.5 \textwidth]{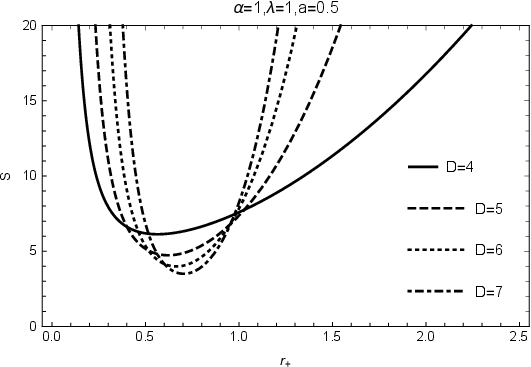}
			\includegraphics[width=0.5 \textwidth]{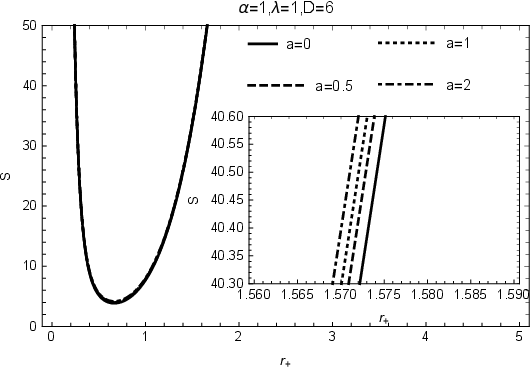}
		\end{tabular}
		\caption{The  entropy $S$ in terms of event horizon $r_{+}$ for different values of $D$ and $a$.} 	\label{FIG2}
	\end{figure}
	
	We draw the corrected entropy versus the event horizon radius for different parameters in Figs.\ref{FIG1} and \ref{FIG2}.  As shown in Fig.\ref{FIG1},  the presence of leading order correction leads to an increase in entropy for small values of the event horizon radius. However, the corrected entropy gradually decreases and recovers to the original entropy when with the increase of the event horizon radius. This means that the equilibrium of the small black hole is unstable due to $\Delta S >0$ when the black hole is regarded as an isolated system. The right figure in Fig. \ref{FIG1} shows that the inverse correction term has a significant influence on the entropy for a small black hole. In fact, compared to the large black hole, the thermal fluctuation has a greater impact on the small black hole. We also show the effect of spacetime dimensionality on the corrected entropy in the left figure of Fig.\ref{FIG2}. We find that the change of corrected entropy is not only fast but also large in high-dimensional spacetime. So one can easily see that for a small or large black hole, the higher the dimension, the larger the corrected entropy, whereas the middle black hole is not the case. We also obtain from the left figure in Fig. \ref{FIG2} that the STVG parameter $a$ leads to a slight increase in corrected entropy.
	
	We can calculate the  Helmholtz free energy using the corrected entropy and temperature  as
	\begin{eqnarray}\label{Q16}
	\begin{aligned}
	F&=-\int S d T = \frac{(D-3)\sqrt{A^{2}-4 G B^{2}}(A-\sqrt{A^{2}-4 G B^{2}})}{8 G B^{2}\pi}  \\ & \times \bigg(\frac{-4 r_{+}^{D-1} \lambda}{(D-1)\Omega_{D-2}} + \frac{r_{+}^{D-3} \Omega_{D-2}}{4(D-3)}+ \frac{\alpha}{r} \big( D-4 +  \text{ln}\bigg[ \frac{(D-3)^{2} \sqrt{A^{2}-4 G B^{2}}(A-\sqrt{A^{2}-4 G B^{2}})^{2} r_{+}^{D-4} \Omega_{D-2}}{256 G^{2} B^{4} \pi^{2}}\bigg]\big)\bigg).
	\end{aligned}
	\end{eqnarray}
	\begin{figure}[t]
		\centering
		\begin{tabular}{cc}
			\includegraphics[width=0.5 \textwidth]{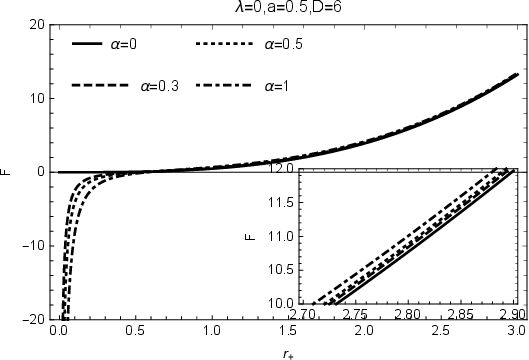}
			\includegraphics[width=0.5 \textwidth]{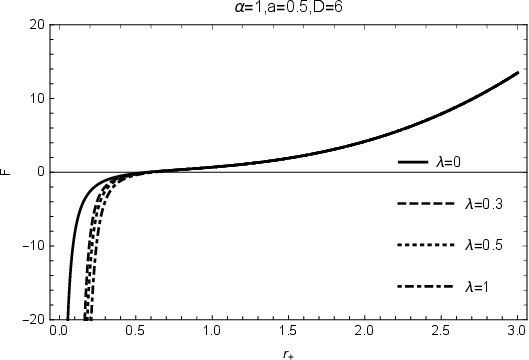}
		\end{tabular}
		\caption{The Helmholtz free energy $F$ in terms of event horizon $r_{+}$ for different values of $\alpha$ and $\lambda$.}	\label{FIG3}
	\end{figure}
	
	\begin{figure}[t]
		\centering
		\begin{tabular}{cc}
			\includegraphics[width=0.5 \textwidth]{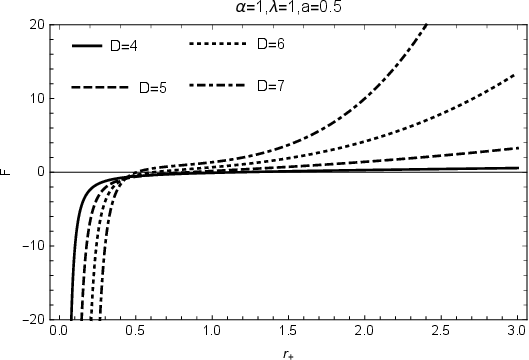}
			\includegraphics[width=0.5 \textwidth]{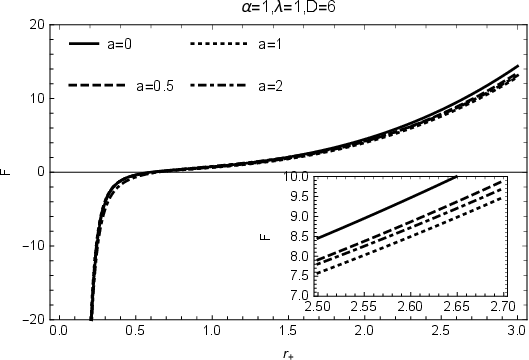}
		\end{tabular}
		\caption{The Helmholtz free energy $F$ in terms of event horizon $r_{+}$ for different values of $D$ and $a$.} 	\label{FIG4}
	\end{figure}
	
	In order to have a better understanding of the corrected Helmholtz free energy, we plot the Helmholtz free energy in terms of the event horizon for the different parameters $\alpha,\lambda, D, a$ in Figs.\ref{FIG3} and  \ref {FIG4}. In Fig.\ref{FIG3}, we can find that the Helmholtz free energy without any corrections is a function that increases monotonically and keeps positive. It is worth noting that the Helmholtz free energy becomes negative for a small black hole under the thermal fluctuation but returns to positive with the increase of event horizon radius.  In contrast to the case of the small black hole, the presence of logarithmic correction term increases the Helmholtz free energy for the larger black hole. We can conclude that thermal fluctuation causes small black holes to be more stable. In addition, we also obtain from the left in Fig.\ref {FIG4} that the impact of spacetime dimension on the modified Helmholtz free energy is similar to that of logarithmic correction. We can see the effect of parameter $a$ on the corrected Helmholtz free energy in the right figure of Fig.\ref {FIG4}. It is clear that the parameter $a$ decreases the corrected Helmholtz free energy.
	
	The internal energy as one of the thermodynamic quantities has the thermodynamics relationship $ E=F+TS$, i.e.,
	\begin{eqnarray}\label{Q17}
	\begin{aligned}
	E&=-\frac{1}{32 \pi (D-1) G B^{2} \Omega_{D-2}}\bigg(4GB^{2}+A(\sqrt{A^{2}-4GB^{2}}-A)r_{+}^{-D-3}\bigg) r_{+}^{-D-3} \\
	& \times  \bigg(16(D-3)(D-2)r_{+}^{4}\lambda+(D-1)r_{+}^{D}\Omega_{D-2} \times \big(4(D-4)(D-3)r_{+}^{2}\alpha+(D-2)r_{+}^{D}\Omega_{D-2}\big)\bigg).
	\end{aligned}
	\end{eqnarray}
	
	\begin{figure}[t]
		\centering
		\begin{tabular}{cc}
			\includegraphics[width=0.5 \textwidth]{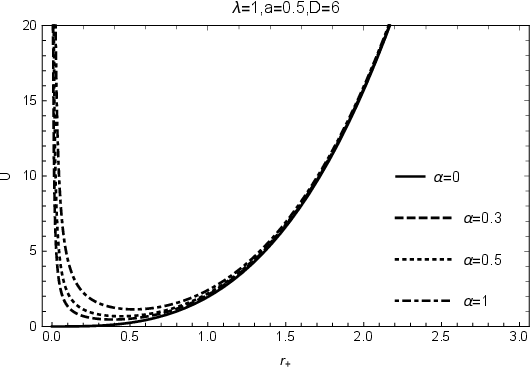}
			\includegraphics[width=0.5 \textwidth]{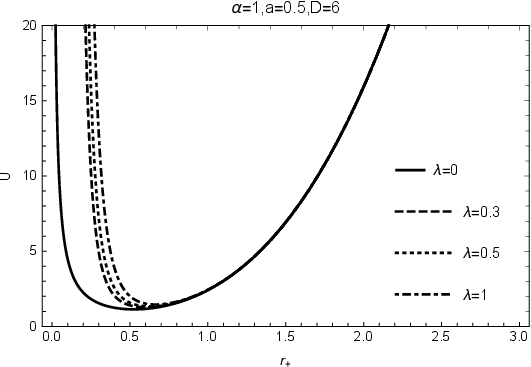}
		\end{tabular}
		\caption{The internal energy $E$  in terms of event horizon $r_{+}$ for different values of $\alpha$ and $\lambda$.}	\label{FIG5}
	\end{figure}
	
	\begin{figure}[t]
		\centering
		\begin{tabular}{cc}
			\includegraphics[width=0.5 \textwidth]{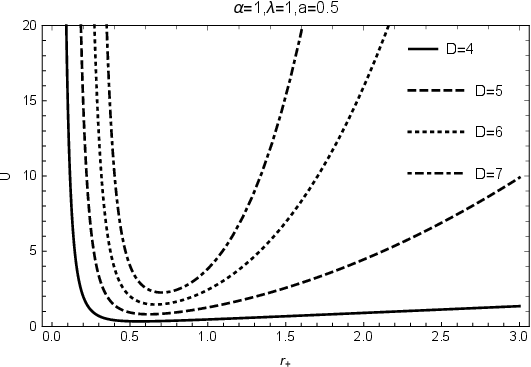}
			\includegraphics[width=0.5 \textwidth]{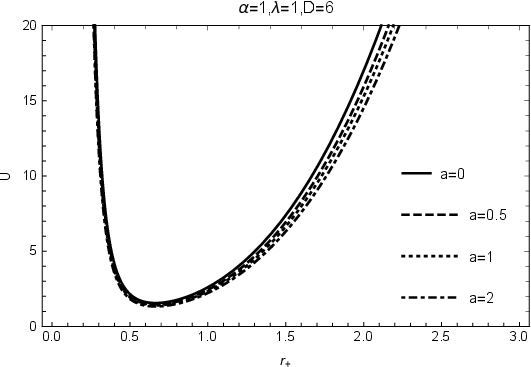}
		\end{tabular}
		\caption{The internal energy $E$ in terms of event horizon $r_{+}$ for different values of $D$ and $a$.} 	\label{FIG6}
	\end{figure}
	
	Figs.\ref{FIG5} and \ref{FIG6} present the behavior of corrected internal energy with increasing the event horizon radius for the different parameters $\alpha,\lambda, D, a$. As it is shown in Fig.\ref{FIG5}, the internal energy has a positive asymptotic value under thermal fluctuation for a small black hole whereas we can neglect the effect of thermal fluctuation when we increase the event horizon radius. We can see clearly that the higher the dimensionality of the black hole, the larger the corrected internal energy. However, the corrected internal energy decreases with the increase of the STVG parameter.

	Next, we investigate the heat capacity of black hole, which can be written as $C=(dU/ dT)_{V}=\frac{(d U/ dr)}{ (dT/dr)}$ using Eqs. (\ref{Q11}) and (\ref{Q17}), concretely
	\begin{eqnarray}\label{Q18}
	\begin{aligned}
	C=(D-4)\alpha+\frac{4(D-2)r_{+}^{D-2}\lambda}{\Omega_{D-2}}-\frac{1}{4}(D-2)r_{+}^{D-2}\Omega_{D-2}.
	\end{aligned}
	\end{eqnarray}
	\begin{figure}[t]
		\centering
		\begin{tabular}{cc}
			\includegraphics[width=0.5 \textwidth]{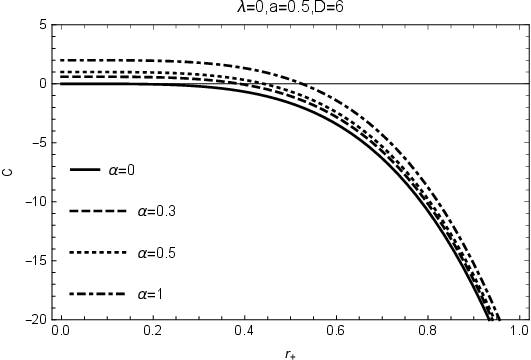}
			\includegraphics[width=0.5 \textwidth]{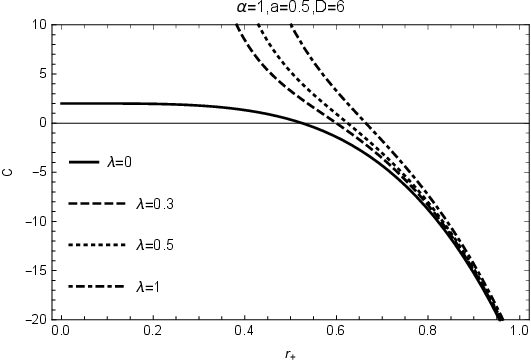}
		\end{tabular}
		\caption{The heat capacity $C$  in terms of event horizon $r_{+}$ for different values of $\alpha$ and $\lambda$.}	\label{FIG7}
	\end{figure}
	
	\begin{figure}[t]
		\centering
		\includegraphics[width=0.5 \textwidth]{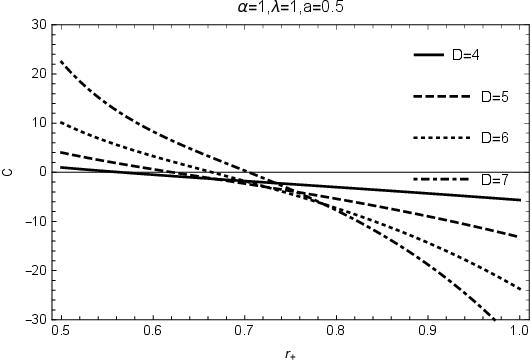}
		\caption{The heat capacity $C$ in terms of event horizon $r_{+}$ for different values of $D$.} 	\label{FIG8}
	\end{figure}
	We draw the behavior of heat capacity by figures of Figs.\ref{FIG7} and \ref{FIG8}. In  Fig.\ref{FIG7}, we observe that without any thermal fluctuation, the heat capacity is negative and thus black hole is thermodynamically unstable. The existence of thermal fluctuations causes small black holes to have positive heat capacity and thus there is a phase transition that shows the transition of the system from stable to unstable. Moreover, the critical point gradually moves to the right when we increase the correction coefficients $\alpha,\lambda$. From Fig.\ref{FIG8}, we can see that  the phase transition occurs at a larger event horizon radius if spacetime dimensionality D increases. It is worth mentioning that the heat capacity of a high-dimensional Schwarzschild black hole in STVG  recovers to that of  Schwarzschild-Tangherlini black hole. That is to say, the STVG parameter does not affect the stability conditions of black holes.

	\section{Weak deflection angle} \label{sec4}
	
	In this section, we would like to obtain the deflection angle in weak field limit using Gauss-Bonnet theorem.  For equatorial plane $\theta =\frac{\pi}{2}$ and null geodesic $ds^{2}=0$, the corresponding  optical metric of  a high-dimensional Schwarzschild black hole in STVG has the  following form
	\begin{equation}
	\label{Q19}
	dt^{2}=\frac{1}{f^{2}(r)}dr^{2}+\frac{r^{2}}{f(r)}d\varphi^{2}.
	\end{equation}
	
	Afterwards, we can rewrite the optical metric using the  coordinate transformation $dr_{*}=\frac{1}{f(r)}dr$ as
	\begin{equation}
	\label{Q20}
	dt^{2}= dr_{*}^{2}+ \tilde{f}^{2}(r_{*})d\varphi^{2},
	\end{equation}
	where $\tilde{f}(r_{*})\equiv\sqrt{\frac{r^{2}}{f(r)}}$.
	
	We obtain the Gaussian optical curvature  as following \cite{Javed:2020lsg}
	\begin{equation}
	\label{Q21}
	\begin{aligned}
	K&=\frac{RicciScalar}{\text{2}} =\frac{1}{4}(D-3)r^{1-4D}\bigg(4(D-2)G^{2}q^{4}r^{9}-2(D-2)Mr^{3D} \\
	& -6(D-2)Gq^{2}r^{6+D}+\big((D-1)M^{2}+4(2D-5)Gq^{2}\big)r^{3+2D}\bigg).
	\end{aligned}
	\end{equation}
	
	Now, we can calculate the deflection angle utilizing Gauss-Bonnet theorem \cite{Gibbons:2008rj}.   The domain $\cal{D}$ is  deemed to be a subset of a compact,  oriented surface,  with Gaussian optical curvature $K$ and Euler characteristic number  $\chi(\cal{D})$ and $\partial \cal{D}$ is the piecewise smooth boundary of domain $\cal{D}$ with geodesic curvature $\kappa$. We consider  $\alpha_{i}$ to be the  $i^{th}$ exterior angle. The Gauss-Bonnet theorem is that
	\begin{equation}
	\label{Q22}
	\int\int_{\cal{D}}K\text{d}S+\int_{\partial \cal{D}}\kappa \text{d} t+\sum_{i}\alpha_{i}=2\pi\chi(\cal{D}),
	\end{equation}
	where $dS$ stands for the surface element. In addition, the geodesic curvature $\kappa$ along a smooth curve $\gamma$ is written as $\kappa=g(\Delta_{\dot{\gamma} }\dot{\gamma},\ddot{\gamma})$ where $\ddot{\gamma}$ denotes unit acceleration vector.  We consider that  $\cal{D}$ is bounded by the geodesics $\gamma_c$ and geodesic $\gamma_R$ where $\gamma_R$ is  considered to be perpendicular to $\gamma_c$ at the source $ S$ and the observer $O$, so $\kappa (\gamma_{c})=0$   by definition.  Then $\sum_{i}\alpha_{i}=\alpha_{S}+\alpha_{O}$ as well as $\chi(\cal{D})=\text{1}$. Eq.(\ref{Q22}) reduces to
	\begin{equation}
	\label{Q23}
	\int\int_{\cal{D}}K\text{d}S+\int_{\gamma_{R}}\kappa (\gamma_{R})\text{d} t =\pi.
	\end{equation}
	
	Utilizing the definition of  geodesic curvature, the radial part of $\kappa (\gamma_{p})$ can be expressed as
	\begin{equation}
	\label{Q24}
	\kappa (\gamma_{p})= (\Delta_{\dot{\gamma_{p}} }\dot{\gamma_{p}})^{r}=\dot{\gamma}_{R}^{\phi}(\partial_{\phi}\dot{\gamma}_{R}^{r})+\Gamma_{\phi\phi}^{r}(\dot{\gamma}_{R}^{\phi})^{2},
	\end{equation}
	where $\dot{\gamma}_{R}$ represents the  tangent vector  of  geodesics $\gamma_R$ and $\Gamma_{\phi\phi}^{r} $  is the  Christoffel symbol.  When we consider $ \gamma_R:=R=const $, the first term on the right side of the above equation equals zero and the second term is $\frac{1}{R}$. So $\kappa (\gamma_{R})$ reduces to $\frac{1}{R}$.
	We can make a change of variables $dt$ using the relevant optical metric (\ref{Q20}), which can be rewritten as $dt=R d\varphi$.
	
	Eq.(\ref{Q23})  becomes
	\begin{equation}
	\label{Q25}
	\int\int_{\cal{D}}K\text{d}S+\int_{0}^{\pi+\alpha} \text{d}\varphi =\pi.
	\end{equation}
	
	Finally, we obtain the deflection angle \cite{Gibbons:2008rj}
	\begin{equation}
	\label{Q26}
	\hat\alpha=-\int_{0}^{\pi}\int_{b/ \sin \phi}^{\infty}K dS.
	\end{equation}
	
	Now, we can calculate the   deflection angle of  a high-dimensional Schwarzschild black hole in STVG for the different  spacetime dimensionality. As an example, we calculate the  deflection angle when $D=4,5,6,7$
	\begin{equation}
	\label{Q27}
	\begin{aligned}
	\hat\alpha_{D=4}&= \frac{2m}{b}-\frac{3 m^{2}\pi}{16b^{2}}-\frac{3G\pi q^{2}}{4b^{2}}+\frac{4Gmq^{2}}{3b^{3}} +O(\frac{q^{4}}{b^{4}}), \\
	\hat\alpha_{D=5}&=\frac{3m\pi}{4b^{2}}-\frac{3m^{2}\pi}{16b^{4}}-\frac{15G\pi q^{2}}{16b^{4}}+\frac{15Gm\pi q^{2}}{32b^{6}}+O(\frac{q^{4}}{b^{8}}), \\
	\hat\alpha_{D=6}&=\frac{8m}{3b^{3}}-\frac{25m^{2}\pi}{128b^{6}}-\frac{35G\pi q^{2}}{32b^{6}}+\frac{512Gmq^{2}}{315b^{9}}
	+O(\frac{q^{4}}{b^{12}}), \\
	\hat\alpha_{D=7}&=\frac{15\pi m}{16b^{4}}-\frac{105m^{2}\pi}{512b^{8}}-\frac{315G\pi q^{2}}{256b^{8}}+\frac{1155Gm\pi q^{2}}{2048b^{12}}
	+O(\frac{q^{4}}{b^{16}}), \\
	\end{aligned}
	\end{equation}
	
	\begin{figure}[t]
		\centering
		\begin{tabular}{cc}
			\includegraphics[width=0.5 \textwidth]{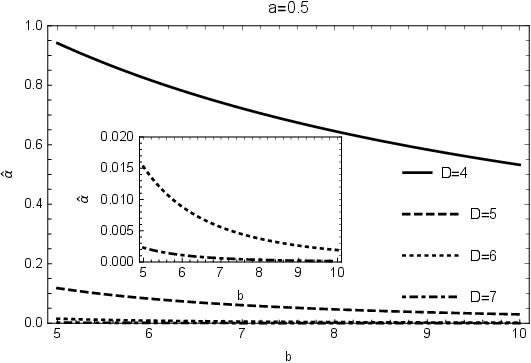}
			\includegraphics[width=0.5 \textwidth]{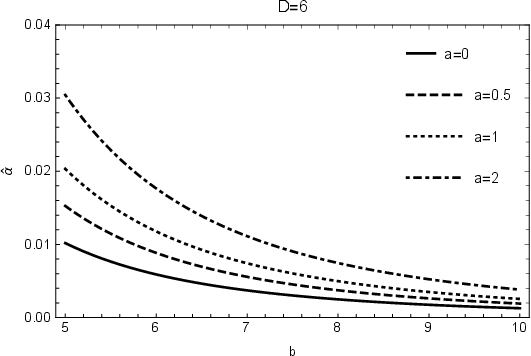}
		\end{tabular}
		\caption{The deflection angle  $\hat\alpha$  in terms of impact parameter $b$ for different values of $D$ and $a$.}	\label{FIG9}
	\end{figure}
	
	We draw the behavior of the deflection angle with respect to the impact parameter for different values of $D$ and $a$ in Fig.\ref{FIG9}. It is clear that the higher the  black hole dimension, the smaller the deflection angle. However, the STVG parameter has an  increasing effect on the deflection angle, i.e.,   a high-dimensional Schwarzschild black hole in STVG leads to a larger deflection angle than a Schwarzschild-Tangherlini black hole.
	
	\section{Greybody factor}\label{sec5}
	
	In this section, we study the bounds on greybody factors for the massless scalar field. The massless scalar field $\Phi$ is represented by the Klein-Gordon equation \cite{Berti:2009kk}
	\begin{equation}
	\label{Q28}
	\frac{1}{\sqrt{-g}}\partial_{\mu}(\sqrt{-g}g^{\mu\nu}\partial_{\nu})\Phi=0,
	\end{equation}
	where $g$ is the determinant of the metric tensor. In order to separate radial and angular variables, we have  an ansatz $\Phi=e^{-i\omega t} Y_{lm}(\Omega)\Psi(r)$ and make a change $dr_{*}=\frac{dr}{f(r)}$. Substituting the above definitions and metric function Eq. (\ref{Q6}) into Eq. (\ref{Q28}), we obtain a Schr$\ddot{\text{o}}$dinger-like wave expression
	\begin{equation}
	\label{Q29}
	\frac{d^{2}\Psi(r)}{d^{2}r_{*}}+[\omega^{2}-V_{eff}(r)]\Psi(r)=0,
	\end{equation}
	in which $ \omega$ donates frequency, $l$ and $m$ are the azimuthal quantum number and the spherical harmonic index, respectively.
	
	The effective potential $V_{eff}(r)$ can be written as
	\begin{equation}
	\label{Q30}
	V_{eff}(r)=f(r)\bigg[\frac{l(D+l-3)}{r^{2}}+\frac{(D-2)(D-4)f(r)}{4r^{2}}+\frac{(D-2)f'(r)}{2r}\bigg].
	\end{equation}
	\begin{figure}[t]
		\centering
		\begin{tabular}{cc}
			\includegraphics[width=0.5 \textwidth]{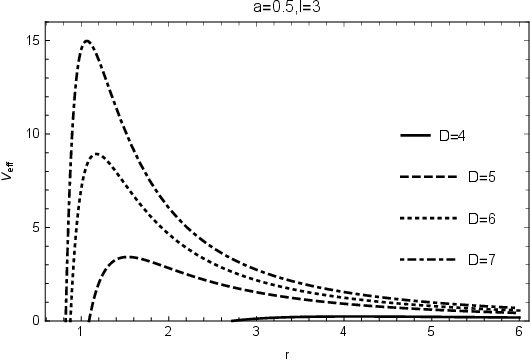}
			\includegraphics[width=0.5 \textwidth]{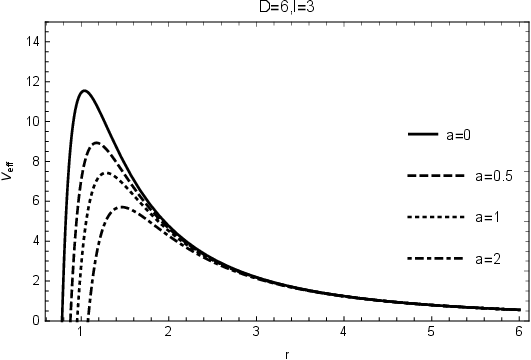}
		\end{tabular}
		\caption{The behavior of  effective potential  $V_{eff}$  for different values of $D$ and $a$.}	\label{FIG10}
	\end{figure}
	
	To better understand the effect of the dimensionality of the spacetime and STVG parameter on the effective potential, we visualize the effective potential with respect to the  black hole radius for different values of $D$ and $a$ in Fig.\ref{FIG10}. Obviously,  the dimensionality of the spacetime causes an increase in the effective potential whereas the STVG parameter has the opposite effect.  We can expect the behavior of greybody factors from the effective potential.
	
	The bounds on greybody factors can be expressed as \cite{Boonserm:2008zg}
	\begin{equation}
	\label{Q31}
	T\geq\text{sech}^{2}\big[\int_{-\infty}^{\infty} \frac{\sqrt{(h')^{2}+(\omega^{2}-V_{eff}-h^{2})^{2}}}{ 2h}dr_{*}\big],
	\end{equation}
	where $ h\equiv h(r_*)$ and $h(r_*)>0$.  $h$ is an arbitrary function and satisfies $h(-\infty)=h(\infty)=\omega$ and  there are two particular functional forms of $h$ considered in Ref.\cite{Boonserm:2008zg}.  Here we only consider the case $h=\omega$. Thus Eq.(\ref{Q31}) is rewritten as
	\begin{equation}
	\label{Q32}
	T\geq\text{sech}^{2}\big[\frac{1}{2\omega}\int_{r_{+}}^{\infty}\frac{V_{eff}}{f(r)}dr\big].
	\end{equation}
	
	After expanding the integral, we obtain the lower bound on the greybody factors
	\begin{equation}
	\label{Q33}
	\begin{aligned}
	T&\geq\text{sech}^{2}\bigg[-\frac{1}{2\omega}\bigg((-8+2D+D^{2}-12l+4lD+4l^{2})(\frac{1}{4r_{+}})\\
	&-\frac{(-2+D)B^{2}(-16+3D)Gm^{2}}{4(2D-5)}r^{5-2D}_{+}+\frac{(D-10)Am}{4}r^{2-D}_{+}
	\bigg)
	\bigg].
	\end{aligned}
	\end{equation}
	
	\begin{figure}[t]
		\centering
		\begin{tabular}{cc}
			\includegraphics[width=0.5 \textwidth]{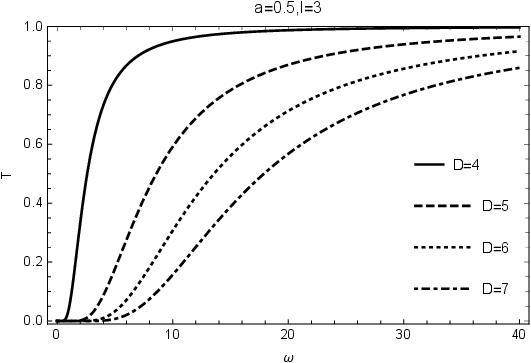}
			\includegraphics[width=0.5 \textwidth]{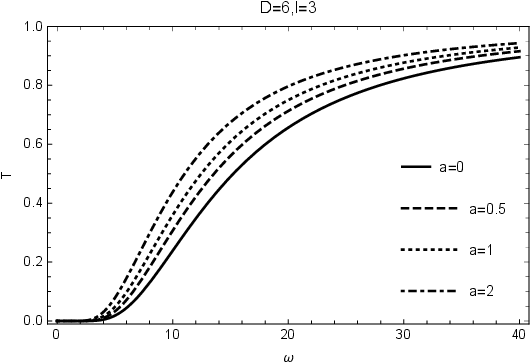}
		\end{tabular}
		\caption{The greybody factor $T$  in terms of frequency  $\omega$ for different values of $D$ and $a$.}	\label{FIG11}
	\end{figure}
	Fig.\ref{FIG11} demonstrates the behavior of the greybody factor for the high-dimensional Schwarzschild black hole in STVG. We observe that the greybody factor reduces with the increase of dimension from the left panel. That is to say, the greybody factor is suppressed in high-dimensional spacetime. It indicates that less massless scalar particles  pass through the  potential barrier and reach to spatial infinity in a higher dimensional black hole.  Additionally, we observe that as the STVG parameter  $a$ increases, the greybody factor increases. That is, the STVG parameter  makes the  gravitational  potential transparent.

	\section{Conclusion}\label{sec6}
	In this paper, we analyzed  thermal fluctuation, weak deflection angle and greybody factor for a high-dimensional Schwarzschild black hole in STVG.
	
	First, we evaluated the influence of the logarithmic and higher-order corrections of the entropy on the  Helmholtz free energy, internal energy and heat capacity and made a comparison to corrected and uncorrected thermodynamic properties. Overall, the corrected entropy as a consequence of thermal fluctuation presents the trend of decreasing first and then increasing, and the impact of thermal fluctuation is significant for a small black hole.  Due to the effect of the dimensionality of spacetime, the curve of modified entropy has different intersections. This causes that for a small-size or large-size black hole,  the corrected entropy increases with the spacetime dimensionality increases, whereas the middle black hole is not the case.  The existence of the STVG parameter leads to a slight increase in corrected entropy.   The black hole with small values of event horizon radius possesses the negative Helmholtz free energy because of the thermal fluctuation. The Helmholtz free energy increases monotonically with increasing values of the parameters $D$ and $a$  for a small-size black hole. For a larger black hole, the parameters $D$ and $a$ have the opposite effects on  Helmholtz free energy. The internal energy remains positive and its behavior is similar to corrected entropy.  The internal energy increases with the increase of dimensions, while it decreases as the STVG parameter increases.  In addition, we found that thermal fluctuation makes the small-size black hole more stable from the analysis of  Helmholtz free energy and heat capacity in all dimensional cases and the heat capacity is independent of the STVG parameter.
	
	Second, we calculated the weak deflection angle with Gauss-Bonnet theorem. We have shown the expression of weak deflection angle for $D=4,5,6,7$. We have pointed out that in the higher dimensional spacetime the weak deflection angle gets weaker but the presence of the STVG parameter results in the increase of deflection angle.
	
	Finally, we computed the greybody factors of the massless scalar field and then analyzed the effect of the spacetime dimensionality and STVG parameter  on greybody factors. We found that the  4-dimensional black hole has the largest values of greybody factors whereas the 7-dimensional black hole possesses the smallest values.  Moreover, we have seen that when the STVG parameter  increases, the greybody factor increases. We got the fact that the more radiation can reach spatial infinity in 4-dimensional black hole with the larger value of STVG parameter.
	
\section*{Declaration of Competing Interest}
The authors declare that they have no known competing financial interests or personal relationships that could have appeared to influence the work reported in this paper.

\end{document}